\documentclass[aps,prd,twocolumn,nofootinbib,amsmath,amssymb]{revtex4-1}
\usepackage{silence} 
\usepackage{CJK}                     
\usepackage[dvips]{graphicx}         
\usepackage{mathptmx}                
\usepackage{bm}
\usepackage{physics}
\usepackage{dcolumn}                 
\usepackage{multirow}                
\usepackage{xcolor}
\usepackage{hyperref}
\hypersetup{
breaklinks=true,colorlinks=true,
linkcolor=blue,citecolor=blue,filecolor=magenta,urlcolor=black}
\long\def\OFF#1{}

\def\be{\begin{equation}} \def\ee{\end{equation}}
\def\bal#1\eal{\begin{align}#1\end{align}}
\def\non{\nonumber}
\def\xv{\bm{x}} \def\pv{\bm{p}}
\def\et{\mathbf{e}}

\def\eps{\varepsilon}
\def\phi{\varphi}
\def\la{\lambda}
\def\al{\alpha}
\def\om{\omega}
\def\ms{\,M_\odot}
\def\mev{\,\text{MeV}}

\def\fm3{\,\text{fm}^{-3}}
\def\mfm{\,\text{MeV}\,\text{fm}^{-3}}
\def\mfms{\;\text{MeV}\,\text{fm}^{-2}}
\def\xp{x_p}
\def\khz{\,\text{kHz}}
\def\deps{\delta\epsilon}
\def\taues{\tau^\text{ES}}
\def\immf{\text{Im}\,\Omega_f}
\def\remf{\text{Re}\,\Omega_f}

\begin{document}

\title{$f$-mode oscillations of hybrid stars with pasta construction}

\begin{CJK*}{UTF8}{gbsn}

\author{Zi-Yue Zheng (郑子岳)$^{1}$}
\author{Jin-Biao Wei (魏金标)$^{2}$}
\author{\hbox{Huan Chen (陈欢)}$^{2}$}\email{Email:huanchen@cug.edu.cn}
\author{Xiao-Ping Zheng (郑小平)$^{1}$}\email{Email:zhxp@ccnu.edu.cn}
\author{G. F. Burgio$^{3}$}
\author{H.-J. Schulze$^{3}$}

\affiliation{
$^{1}$Institute of Astrophysics, Central China Normal University,
Luoyu Road 152, 430079 Wuhan, China\\
$^{2}$School of Mathematics and Physics, China University of Geosciences,
Lumo Road 388, 430074 Wuhan, China\\
$^{3}$INFN Sezione di Catania, Dipartimento di Fisica,
Universit\'a di Catania, Via Santa Sofia 64, 95123 Catania, Italy
}

\begin{abstract}
We investigate nonradial $f$-mode oscillations
of hybrid neutron stars in full general relativity,
employing hybrid equations of state
describing a nuclear outer core and a pasta-phase transition
to a quark-matter core.
The validity of various universal relations is confirmed
for those stars.
Prospects of observations are also discussed.
\end{abstract}

\maketitle
\end{CJK*}

\section{Introduction}

Neutron stars (NSs) are the densest stars observed in the Universe
and provide natural laboratories for studying high-density nuclear matter (NM).
Due to the core density of a NS predicted to be several times
the nuclear saturation density $\rho_0 \approx 0.16\fm3$,
a phase transition to deconfined quark matter (QM)
may take place in its interiors
\cite{Bauswein19,Annala20}.
Hence, NSs can be used to constrain theories concerning high-density QM
and NM-QM phase transitions, for which the
Maxwell construction (MC) and the
Gibbs construction (GC) are commonly used.

It is noteworthy that only bulk-matter contributions are included in GC and MC,
where finite-size effects depending on a surface tension $\sigma$ are neglected.
The MC corresponds to a extremely large surface tension
and leads to a bulk separation of quark and hadron phases
\cite{Bhattacharyya10,Han19},
whereas the GC corresponding to a vanishing surface tension
determines a range of baryon number densities where the two phases coexist
\cite{Glendenning92,Schertler00,Yang08,Chen13}.
If a moderate surface tension is employed,
a hadron-quark mixed phase (MP) with pasta structures is expected to occur
\cite{Heiselberg93,Endo06,Maruyama07,Yasutake14,Wu19,Xia20,Ju21}.
Thus the surface tension $\sigma$ plays  a key role
in determining the various structures in the hadron-quark MP.
It is therefore interesting to examine what pasta construction
can be realized in NSs and how it can influence the NS properties.

After the first direct observation of gravitational waves (GWs)
from a binary black hole (BH) merger \cite{Abbott16},
more and more GW signals were detected,
including binary NS systems \cite{Abbott17a,Abbott20a}
and NS-BH systems \cite{Abbott19a}.
These GW observations have opened a new window for studying
the internal structure of NSs,
especially in helping us to understand the high-density NM \cite{Abbott18}
and the astrophysics processes under extreme conditions \cite{Abbott17b}.
Recently, due to GW observations from possibly the most massive NSs
ever detected,
hybrid stars (HSs) have once again become popular subjects
\cite{Tan20,Ferreira20,Dexheimer21,Demircik21,Blaschke21}.

NSs are expected to emit various electromagnetic signals and GWs
during specific astrophysical processes.
Stellar oscillations, which are a sensitive probe of NS internal composition,
may provide a tool for addressing the problem of high-density NM.
Especially, the ($l\geq2$) nonradial oscillation (NRO) of a star
can carry information about the core of a NS through GW emission
\cite{Thorne67},
which may occur during a supernova explosion \cite{Radice19},
or during the post-merger phase of a binary NS
\cite{Kokkotas01,Stergioulas11,Vretinaris20,Soultanis22},
or in an isolated perturbed NS \cite{Doneva13}.
For a non-rotating star, NRO can be divided into various modes:
for fluid modes,
these are $p$ (pressure)-mode, $f$ (fundamental)-mode, and $g$ (gravity)-mode,
which indicate the various dominant restoring forces for the perturbations.
The buoyancy acts as the $g$-mode restoring force
to bring perturbed fluid elements back into equilibrium,
while the restoring force of $f$-mode and $p$-mode is pressure.


Since the $g$-mode eigenfrequencies are relatively small
and provide us with an appropriate observable,
many works use them as probes for new degrees of freedom within a NS,
such as deconfined QM
\cite{Weiwei20,Bai21,Jaikumar21,Constantinou21,Zhao22a,Zheng23},
hyperons \cite{Dommes16,Tran23},
and superfluidity \cite{Kantor14,Dommes16,Hang16,Rau18}.
However, the $p$-mode eigenfrequency is too high to be observed,
far beyond the sensitivity range of the next-generation GW detectors
\cite{Kokkotas99,Kunjipurayil22}.
The $f$-mode eigenfrequencies lie between that of $g$ and $p$-mode,
with frequency
$f_f = \Re\om_f / 2\pi \sim 1.3-3 \khz$.
It was argued that the excitation of the $f$-modes can be enhanced
by the spin and eccentricity during the inspiral phase of NS mergers
\cite{Chirenti17,Steinhoff21}.
The $f$-mode frequency is believed to be related to the mean density of the NSs
\cite{Andersson98,Kokkotas99,Pradhan22},
the pressure of matter in $\beta$ equilibrium at densities
equal to or slightly higher than the nuclear saturation density
\cite{Kunjipurayil22},
and the tidal deformability during the inspiral phase of NS mergers
\cite{Chan14,Hinderer16,Andersson21}.

From the asteroseismology of NSs, a set of
so-called universal relations (URs) has been proposed.
This method is expected to help estimating macroscopic quantities of NSs,
such as the mass, radius, or tidal deformability,
by correlating them
with measured frequency and damping time of a particular oscillation mode.
Among the many studies in the literature investigating the $f$-mode oscillation,
\cite{Kokkotas99,Andersson98} first related the NS global properties
with the frequency and damping time of that mode.
Since this pioneering work,
many alternative URs have been proposed,
correlating the dimensionless frequency $\Omega_f \equiv M\om_f$
with other NS properties
such as the dimensionless moment of inertia $\bar{I} \equiv I/M^3$
\cite{Lau09,Chirenti15}
and the dimensionless tidal deformability $\Lambda$ \cite{Chan14,Sotani21}.
These URs have been examined in NSs with various configurations,
including purely hadronic stars \cite{Sotani21},
hyperonic stars \cite{Pradhan22},
hybrid stars with quark-hadron crossover \cite{Sotani23}
or MC \cite{Zhao22b},
and strange quark stars \cite{Zhao22b}.
Thus, it is natural to investigate whether NSs with hadron-quark pasta structure
also follow the same URs.

In this article,
we adopt the Brueckner-Hartree-Fock (BHF) theory for NM
\cite{Jeukenne76,Baldo99},
which is based on realistic two-nucleon forces
supplemented with compatible microscopic three-nucleon forces
\cite{Grange89,Zuo02,Li08a,Li08b},
and features reasonable properties at nuclear saturation density
in agreement with experiments
\cite{Li08a,Li08b,Kohno13,Fukukawa15,Wei20,Burgio21}.
Moreover, the BHF approach is also compatible with observable constraints
obtained from the analysis of the GW170817 event and NICER mission
\cite{Wei20,Burgio21,Wei21}.
For QM we adopt a model in the framework of Dyson-Schwinger equations (DSM)
\cite{Chen11,Chen15},
which provides a continuum approach to quantum chromodynamics (QCD)
that can simultaneously address both confinement
and dynamical chiral symmetry breaking
\cite{Roberts94,Alkofer01}.
We employ the pasta construction between the hadron and deconfined quark phase,
which determines a range where hadron and quark phase coexist.
In this framework,
we study the URs of HSs with pasta construction in full general relativity (GR).

This work is organized as follows.
In Sec.~\ref{s:eos} we review the formalism for the EOSs, i.e.,
the BHF theory for the hadron phase and the DSM for the quark phase.
In Sec.~\ref{s:osc} we introduce the macroscopic features and
the eigenvalue equations for the NROs of NSs.
Numerical results are given in Sec.~\ref{s:res},
and we draw the conclusions in Sec.~\ref{s:end}.
We use natural units $G=c=\hbar=1$ throughout the article.

\section{Equation of state}
\label{s:eos}

\subsection{Nuclear matter}

The essential ingredient in BHF many-body approach
is the in-medium Brueckner reaction matrix $K$,
which is the solution of the Bethe-Goldstone equation
\be
 K(\rho,\xp;E) = V + \Re\sum_{1,2}V
 \frac{\ket{12} Q \bra{12}}{E-e_1-e_2} K(\rho,\xp;E) \:
\label{e:k}
\ee
and
\be
 U_1(\rho,\xp) = \Re\sum_{2<k_F^{(2)}}
 \expval{G(\rho,\xp;e_1+e_2)}{12}_a \:,
\label{eq:uk}
\ee
where $V$ is the nucleon-nucleon interaction potential,
$\xp \equiv \rho_p/\rho$ is the proton fraction,
and $\rho_p$ and $\rho$ are the proton and the total nucleon number densities,
respectively.
$E$ is the starting energy,
$Q$ is the Pauli operator, and
$e_i \equiv k_i^2\!/2m_i + U_i$ is the single-particle energy.
The multi-indices $1,2$ denote in general momentum, isospin, and spin.
The corresponding BHF energy density can then be expressed as
\be
 \eps = \sum_{1<k_F^{(1)}}
 \qty( \frac{k^2}{2m_1} + \frac12 U_1(k) )
\:.
\label{eq:f}
\ee
The chemical potentials of the nucleons can be derived in a consistent way,
\be
 \mu_i = {\frac{\partial\eps}{\partial\rho_i}} \:.
\ee
We impose cold, neutrino-free, charge neutral, and catalyzed matter
consisting of neutrons, protons, and leptons ($e^-,\mu^-$)
in beta equilibrium due to weak interactions.
Finally, the pressure is given by
\be
 p(\eps) = \rho^2 \frac{\partial}{\partial\rho}
 \frac{\eps(\rho_i(\rho))}{\rho}
 = \sum_i \rho_i \mu_i - \eps \:.
\ee

In the BHF approach,
the nucleon-nucleon interaction potential $V$ is the only necessary input.
In this work, we employ the Argonne $V_{18}$ (V18) \cite{Wiringa94}
and Bonn-B (BOB) \cite{Machleidt87,Machleidt89} potentials,
supplemented with compatible microscopic three-body forces
\cite{Grange89,Zuo02,Li08a,Li08b}.
With this common prescription,
the saturation point of symmetric NM
and related properties can be reproduced correctly.
We use the convenient empirical parametrizations
given in Refs.~\cite{Lu19,Wei20}.

The BHF approach provides only the EOS for the bulk-matter core region
without cluster formation,
and therefore has to be combined with a low-density crust EOS.
In this work, we adopt the unified Shen2020 EOS \cite{Shen20},
which is frequently used
for the simulations of core-collapse supernovae and NS mergers.
We also calculated the results using the full Shen2020 EOS for comparison.
Since there are no rigorous theoretical or observational limitations on
how to join the crust and core EOSs,
we opt for the most straightforward and simplest method,
namely a continuous transition
at locations where pressure and energy density coincide.

\subsection{Quark matter}

For the description of deconfined QM,
we adopt the framework of Dyson-Schwinger equations
\cite{Chen11,Chen12},
which are the fundamental equations of motion in continuum quantum field theory.
The fundamental quantity of the DSM is the quark propagator $S(p;\mu)$
at zero temperature and finite chemical potential $\mu$,
which can be expressed as
\bal
 & S(p;\mu)^{-1} = Z_2 \left[ i \bm{\gamma} \cdot \pv
 + i \gamma_4 \tilde{p}_4  + m_q \right] + \Sigma(p;\mu) \:,
 \label{e:dse}
\eal
where $p=(\pv,p_4)$ is the four-momentum,
$\tilde{p}_4 = p_4 + i\mu$,
and $m_q$ is the current mass of the quark $q=u,d,s$.
In this work,
we choose $m_u = m_d = 0$ and $m_s = 115\mev$.
The renormalized self-energy $\Sigma(p;\mu)$ is expressed as
\bal
 \Sigma(p;\mu) = &~Z_1 g^2(\mu) \non\\
 &\times \!\! \int \!\!
 \frac{d^4 q}{(2\pi)^4} D_{\rho\sigma}(p-q;\mu)
 \frac{\la^a}{2} \gamma_\rho S(q;\mu)
 \frac{\la^a}{2} \Gamma_\sigma(q,p;\mu) \:,
 \label{dssigma}
\eal
where $Z_1$ and $Z_2$ are the quark-gluon vertex and
quark wavefunction renormalization constants, respectively.
$D_{\rho\sigma}$ is the full gluon propagator,
$\Gamma_\sigma$ is the full quark-gluon vertex,
$\la^a$ are the Gell-Mann matrices,
and $g(\mu)$ is the density-dependent coupling constant.
To obtain the quark propagator,
the so-called rainbow approximation and a
chemical-potential-modified Gaussian-type effective interaction
\cite{Chen11,Luo19} are adopted.

The quark number density, pressure, and energy density
for each quark flavor at zero temperature can be obtained as
\cite{Chen08,Klahn09}
\bal
 \rho_q(\mu_q) &= 6\int\!\! \frac{d^4p}{(2\pi)^4}
 \mathop{\text{tr}_D} \left[ -\gamma_4 S_q(p;\mu_q) \right] \:,
\label{e:dsrho}
\\
  p_q(\mu_q) &= p_q(\mu_{q,0}) + \int_{\mu_{q,0}}^{\mu_q} d\mu \rho_q(\mu) \:,
\label{e:dsp}
\\
  \eps_q(\mu_q) &= -p_q(\mu_q) + \mu_q \rho_q(\mu_q) \:.
\label{e:dsed}
\eal
The total pressure and energy density are given
by summing contributions from all quark flavors
and those from electrons and muons.
The pressure of QM at zero density is determined by a
phenomenological bag constant \cite{Wei17},
\be
 B_\text{DS} = -\!\!\sum_{q=u,d,s} p_q(\mu_{q,0}) \:,
\label{BDS}
\ee
which is set to $90\mfm$ \cite{Chen12,Chen15,Wei17}.

\subsection{Hadron-quark mixed phase}

It is well known that the surface tension $\sigma$ at the NM-QM interface
plays a crucial role in determining the structure of the MP.
GC and MC correspond to the two extreme cases of
zero and very large surface tension, respectively.
In this work we adopt the energy-minimization method \cite{Wu19,Ju21}
to obtain the pasta structure of the MP,
where the hadronic and quark phases inside a Wigner-Seitz cell
are assumed to be separated by a sharp interface with a finite surface tension.
The geometric configuration of pasta structures
may change from droplet to rod, slab, tube, and bubble
with increasing baryon number density
\cite{Endo06,Maruyama07,Yasutake14,Xia20,Wu19,Ju21,Xia20}.

However, the value of the surface tension is still poorly known so far.
Some works based on, e.g.,
MIT bag model \cite{Berger87,Lugones17,Lugones19},
linear sigma model \cite{Palhares10,Kroff15},
a geometrical approach \cite{Pinto12},
three-flavor Polyakov quark-meson model \cite{Mintz13},
Nambu-Jona-Lasinio (NJL) model \cite{Garcia13,Ke14},
nucleon-meson model \cite{Fraga19},
gave relatively small values of the surface tension
$\sigma \lesssim 30\mfms$.
One work based on the DSM predicts the interface tension at zero temperature
to be $25.4\mfms$ for the hadronization process
and $40.0\mfms$ for the opposite \cite{Gao16}.
On the other hand,
large values of $\sigma$ are also obtained by using
the multiple reflection expansion method in the NJL model
including color superconductivity,
which predicted $\sigma \approx 145-165\mfms$ \cite{Lugones13},
and from the extended quasiparticle model \cite{Wen10}
and color-flavor-locked phase \cite{Alford01}.
Due to this uncertainty,
we treat the surface tension as a free parameter,
comparing results with
$\sigma = 0~\text{(corresponding to GC)},10$, and $30\mfms$.

In our model,
all configurations of HSs with MC are unstable,
due to violation of the stability condition
$\partial M / \partial \eps_c \geq 0$
at the QM onset.

\section{Neutron Stars}
\label{s:osc}

\subsection{Macroscopic features}

With a specified EOS, the mass-radius relation,
one of the most straightforward and simple macroscopic features,
can be obtained by solving the TOV equations.
The Schwarzschild metric for a spherically symmetric system is given by
\be
 ds^2 = e^{\nu(r)}dt^2 - e^{\la(r)}dr^2
 - r^2(d\theta^2+\sin^2\!\theta d\phi^2) \:,
\label{e:ds2}
\ee
where $e^{\nu(r)}$ and $e^{\la(r)}$ are metric functions.
The TOV equations \cite{Oppenheimer39,Tolman39}
for the hydrostatic equilibrium of stars in GR are given by
\bal
 \frac{dp}{dr} &= \frac{(p+\eps)(m+4\pi r^3p)}{r^2(2m/r-1)} \:,
\label{e:dpdr}
\\
 \frac{dm}{dr} &= 4\pi r^2\eps \:, \label{e:dmdr}
\eal
where $p$, $\eps$, and $m$ are the pressure,
the energy density,
and the enclosed gravitational mass, respectively.
The corresponding metric functions are given by
\bal
 e^{\la} &= \frac{1}{1-2m/r} \:,
\\
 \frac{d\nu}{dr} &= -\frac{2}{p+\eps} \frac{dp}{dr} \:.
\label{e:metric}
\eal
One can obtain radius $R$ and mass $M=m(R)$
of a NS for a given central pressure or density
by solving Eqs.~(\ref{e:dpdr},\ref{e:dmdr})
with the boundary condition $p(R)=0$.

GWs from the merger of binary NSs or a BH with a NS,
now provide valuable information on the EOS and internal structure of NSs,
in particular their dimensionless tidal deformability,
defined by the tidal Love number $k_2$ \cite{Hinderer08,Hinderer10},
\be
 \Lambda = \frac23 \frac{k_2}{\beta^5} = \frac{16}{15} \frac{z}{F}
\label{e:tdef}
\ee
with $\beta \equiv M/R$ the compactness of the star and
\bal
 z \equiv &~ (1-2\beta)^2 [2-y_R+2\beta(y_R-1)] \:,
\non\\
 F \equiv &~ 6\beta(2-y_R) + 6\beta^2(5y_R-8) + 4\beta^3(13-11y_R)
\non\\
          &  + 4\beta^4(3y_R-2) + 8\beta^5(1+y_R) + 3z\ln(1-2\beta) \:.
\eal
$y_R=y(R)$ can be obtained by solving the additional differential equation
for $y(r)$ \cite{Lattimer06}
\be
 \frac{dy}{dr} = -\frac{y^2}{r} - \frac{y-6}{r-2 m} - r Q \:,
\ee
where
\be
 Q = 4\pi \frac{(5-y)\eps + (9+y)p + (p+\eps)(dp/d\eps)^{-1}}{1-2m/r}
 - \bigg(\frac{d\nu}{dr}\bigg)^2
\ee
with the boundary condition $y(r=0)=2$.

Moreover, in GR for a uniformly slowly rotating star,
the moment of inertia (MOI) can be expressed as \cite{Wei19}
\be
 I = \frac{w_R R^3}{6+2w_R} \:,
\label{e:moi}
\ee
where $w_R=w(R)$ can be obtained by solving the differential equation
\be
 \frac{dw}{dr} = 4\pi r \frac{(p+\eps)(4+w)}{1-2m/r} - \frac{w}{r}(3+w) \:,
\ee
with the boundary condition $w(0)=0$.

\subsection{Nonradial oscillations}

Thorne et al.\ developed a complete theory for NROs of NSs in GR
\cite{Thorne67}.
In this work,
we study only even-parity perturbations of the Regge-Wheeler metric,
\bal
 ds^2 =& -e^{\nu(r)} \big[
 1 + r^l H_0(r) e^{i\om t} Y_{lm}(\theta,\phi) \big] dt^2
\non\\
       & +e^{\la(r)} \big[
 1 - r^l H_0(r) e^{i\om t} Y_{lm}(\theta,\phi) \big] dr^2
\non\\
 & + \big[ 1 - r^l K(r) e^{i\om t} Y_{lm}(\theta,\phi) \big]
 r^2 ( d\theta^2 + \sin^2\!\theta d\phi^2 )
\non\\
 & -2i\om r^{l+1} H_1(r) e^{i\om t} Y_{lm}(\theta,\phi) dt dr \:,
\eal
where $Y^l_m(\theta,\phi)$ are the usual spherical harmonics,
and $H_0$, $H_1$, and $K$ are metric perturbation functions.
$\om = 2\pi f + i/\tau$ is the complex eigenfrequency of the NRO,
with $\tau$ the damping time of the GW.
The perturbation of the fluid in the star is described
by the Lagrangian displacement vector $\xi^\al$ in terms of
the dimensionless eigenfunctions $W(r)$ and $V(r)$
of the radial and transverse fluid perturbations,
\bal\label{e:xi}
 \xi^r      &= r^{l-1} e^{-\la/2} W Y^l_m \,e^{i\om t} \:,
\non\\
 \xi^\theta &= - r^{l-2} V \partial_\theta Y^l_m \,e^{i\om t} \:,
\non\\
 \xi^\phi   &= - r^{l-2} (\sin\theta)^{-2} V
 \partial_\phi Y^l_m \,e^{i\om t} \:.
\eal
An additional fluid perturbation amplitude $X$,
related to the Lagrangian perturbation of the pressure,
is defined as
\be
 \Delta p = -r^l e^{-\nu/2} X Y_m^l e^{i\om t} \:.
\ee
The full NRO equations governing these perturbation functions
and the metric perturbations inside the star are given by
\cite{Thorne67,Detweiler85,Zhao22b}
\bal\label{e:os1}
 r\frac{dH_1}{dr} &= -\big[ l+1 + 2be^\la + 4\pi r^2 e^\la(p-\eps) \big] H_1
 + e^\la\big[ H_0 + K - 16\pi q V \big] \:,
\\
 r\frac{dK}{dr} &= H_0 + (n+1)H_1
 + \big[ e^\la Q - l - 1 \big] K
 - 8\pi q e^{\la/2} W \:,
\\
 r\frac{dW}{dr} &= r^2 e^{\la/2}
 \bigg[ \frac{H_0}{2} + K + \frac{e^{-\nu/2}}{q c_s^2} X \bigg]
 - (l+1) \big[W + l e^{\la/2} V \big] \:,
\\
 r\frac{dX}{dr} &= - l X + \frac{q}{2} e^{\nu/2} \Bigg\{
 -\frac{4(n+1)e^\la Q}{r^2} V
\non\\&
 + \big[ 3e^\la Q - 1 \big] K + \big[ 1 - e^\la Q \big] H_0
 + \big[ (\om r)^2 e^{-\nu} + n + 1 \big] H_1
\non\\&
 - e^{\la/2} \bigg[ \frac{2\om^2}{e^\nu} + 8\pi q
 - \frac{r^2}{e^{\la/2}} \frac{d}{dr}
 \bigg( \frac{e^{-\la/2}}{r^2} \frac{d\nu}{dr} \bigg) \bigg] W
 \Bigg\} \:,
\label{e:os4}
\eal
where $n \equiv (l-1)(l+2)/2$,
$b \equiv m(r)/r$,
$Q \equiv 4\pi r^2 p + b$,
$q \equiv p+\eps$,
and $c_s$ is the adiabatic sound speed.
In the following,
we adopt the equilibrium sound speed
$c_e = \sqrt{dp/d\eps}$ as an approximation,
defined under the assumption
that reactions are faster than the oscillation.

The perturbation functions $H_0$ and $V$ can be expressed as
\bal
 H_0 &= (n + 2b + Q)^{-1} \Big\{
 \big[ (\om r)^2 e^{-(\nu+\la)} - (n+1)Q \big] H_1
\non\\&\hskip2mm
 + \big[ n - (\om r)^2 e^{-\nu} - Q^2 e^\la + Q \big] K
 + 8\pi r^2 e^{-\nu/2} X
 \Big\} \:,
\\
 V &= \frac{e^{\nu}}{\om^2}
 \bigg[ \frac{e^{-\nu/2}}{q} X - \frac{Q}{r^2} e^{\la/2} W
 - \frac{1}{2} H_0 \bigg]
\:.
\eal

In order to solve the NRO eigen-equations Eq.~(\ref{e:os1}-\ref{e:os4}),
one also needs boundary conditions,
which, in the NS center $r=0$, are given by
\bal
 H_1 &= \frac{l K + 8\pi q W}{n+1} \:,
\non\\
 K &= \pm q \:,
\non\\
 W &= 1 \:,
\non\\
 X &= q e^{\nu/2} \left\{
 \left[ \frac{4\pi}{3} (3p+\eps) - \frac{\om^2}{l e^\nu} \right] W
 + \frac{K}{2} \right\} \:,
\label{e:boundary_c}
\eal
while at the outer surface of the star,
the Lagrangian perturbation of the pressure should vanish,
\be
 X(R) = 0 \:.
\label{e:boundary_out}
\ee

The perturbations of the metric outside the star
with the disappearance of fluid perturbations
can be described with a unique function $Z$,
which satisfies the Zerilli equation
\be
 \frac{d^2 Z}{dr^{*2}} = \left[ V_Z(r) - \om^2 \right] Z \:,
\label{e:zerilli}
\ee
where $r^* \equiv r + 2M\ln(1/2b-1)$  
and now $b=M/r$ with the NS mass $M$.
The metric functions are
\bal
 \binom{K(r)}{H_1(r)} &= \left(\begin{array}{cc}
 g(r) & 1 \\
 h(r) & k(r)
\end{array}\right) \binom{Z(r^*)/r}{dZ(r^*)/dr^*} \:,
\non\\
 g(r) &= \frac{n(n+1)+3nb+6b^2}{(n+3b)} \:,
\non\\
 h(r) &= \frac{n-3nb-3b^2}{(1-2b)(n+3b)} \:,
\non\\
 k(r) &= \frac{dr^*}{dr}(r) = \frac{1}{1-2b} \:,
\label{e:boundary_z1}
\eal
and
\be
 H_0 = \frac{\big[
 (\om r)^2 - (n+1)b \big] H_1 + \left[n(1-2b) - (\om r)^2 + b(1-3b) \right] K}
 {(1-2b)(3b+n)} \:,
\ee
whereas the effective potential is
\be
 V_Z(r) = 2(1-2b) \frac{n^2(n+1)+3n^2b+9nb^2+9b^3}{(n+3b)^2 r^2} \:.
\ee

The asymptotic solution of $Z$ at large $r$ can be decomposed
into incoming and outgoing wave functions as
\bal
 \binom{Z}{dZ/dr^*} =& \left(\begin{array}{cc}
 Z_\text{out} & Z_\text{in} \\
 dZ_\text{out} / dr^* & dZ_\text{in}/dr^*
 \end{array}\right)
 \binom{A_\text{out}}{A_\text{in}} \:,
\eal
where
\bal
 Z_\text{out} =& e^{-i\om r^*} \al \Big[
 1 + \frac{c_1}{\om r} + \frac{c_2}{(\om r)^2} + \mathcal{O}(r^{-3}) \Big] \:,
\non\\
 \frac{dZ_\text{out}}{dr^*} =& -i\om e^{-i\om r^*} \al \Big[
 1 + \frac{c_1}{\om r}
 + \frac{c_2+ic_1(1-2 b)}{(\om r)^2} + \mathcal{O}(r^{-3}) \Big] \:,
\non\\
 c_1 =& -i(n+1) \:,
\non\\
 c_2 =& [-n(n+1)+i(3+6/n)M\om]/2 \:.
\label{e:analytic_z}
\eal
Here $A_\text{out}$ and $A_\text{in}$ are the amplitudes
of outgoing and incoming waves, respectively.
$Z_\text{in}$ is the complex conjugate of $Z_\text{out}$
and $\al$ can be any complex number.
At infinity, the gravitational radiation field should be purely outgoing,
in other words, the amplitude $A_\text{in}$ is zero.

We now briefly introduce our numerical method
for the quadrupole $(l=2)$ $f$-mode oscillations.
First,
starting with two sets of central boundary conditions~(\ref{e:boundary_c})
and integrating Eqs.~(\ref{e:os1}-\ref{e:os4}) from NS center to the surface
one obtains two trial solutions.
These are linearly combined to obtain a solution
satisfying also the outer boundary condition~(\ref{e:boundary_out}).
Outside the star, the boundary values of the Zerilli equation~(\ref{e:zerilli})
at the outer surface of the star are fixed using (\ref{e:boundary_z1})
at $r=R$.
Then the equation is integrated numerically
up to $r=50\;\Re\om^{-1}$,
which is found to be adequate for the asymptotic solution \cite{Detweiler85}.
The complex coefficients $A_\text{in}$ and $A_\text{out}$ are obtained by
matching the analytic expressions~(\ref{e:analytic_z}) for $Z$ and $dZ/dr^*$
with the numerical results.
Finally, we adopt the method of continued fractions to
search the complex eigenfrequencies $\om$ satisfying $A_\text{in}=0$.

We tested our numerical routines
by reproducing the quadrupole $f$-mode complex frequencies
from Ref.~\cite{Sotani01} for polytropic EOSs,
$f$-mode perturbation functions and complex frequencies
from Ref.~\cite{Zhao22b} for SLy4 EOS,
and $p$-mode perturbation functions and complex frequencies
from Ref.~\cite{Kunjipurayil22} for SLy4 EOS.

\subsection{GW strain}

In quadrupole approximation and transverse-traceless (TT) gauge,
the GW strain (metric perturbation) is \cite{Thorne80,Szczepanczyk21}
\be\label{e:h}
 h_{ij}^\text{TT}(t,D) =
 \frac{2}{D} \ddot{Q}_{ij}^\text{TT}(t-D) \:,
\ee
where $i,j=1,2,3$ are the indices in Cartesian coordinates,
$D$ is the distance to the source,
the two dots represent the second time derivative,
and the traceless quadrupole moment is
\be\label{e:Q}
 Q_{ij}^\text{TT}(t) = \int d^3\xv\, \eps(t,\xv)
 \Big( x_i x_j - \frac13 \delta_{ij}|\xv|^2 \Big) \:.
\ee
The metric perturbation tensor can be decomposed as
\be\label{e:h1}
 \mathbf{h}^\text{TT} = h_+ \et_+ + h_\times \et_\times \:,
\ee
where $\et_+$ and $\et_\times$
are the unit tensors of plus and cross polarization.
In the case of a symmetric metric,
the cross polarization $h_\times$ is zero,
$Q_{ij}$ has only diagonal components
$Q_{11} = Q_{22} = -\frac12 Q_{33}$
(for simplicity, we omit TT from $Q_{ij}$),
and the GW strain can be calculated by \cite{Finn90}
\be
 h_+ = \frac{3\sin^2\theta}{2D} \ddot{Q}_{33} \:,
\label{e:hp1}
\ee
where $\theta$ is the inclination angle.
In this work, we choose $\sin\theta=1$.
After some derivation, $\ddot{Q}_{33}$ can be written as \cite{???}
\bal
 \ddot{Q}_{33}
 &= \left| \ddot{Q}_{33}\right| e^{-i\om t}
\non\\
 &= \int dr\, r^2 \sin\theta d\theta d\phi\,
 r^2 \left(\sin^2\theta - \frac13 \right)
 \frac{d^2 \deps}{dt^2} \:,
\eal
and thus
\be
 \left|\ddot{Q}_{33}\right| =
 \frac{4\sqrt{\pi/5}}{3} \om^2 \int dr\, r^4 \deps(r) \:,
\label{e:Q33}
\ee
where $\deps$ is the Eulerian perturbation of the energy density,
given as \cite{Thorne67}
\bal
 \deps = & -(p+\eps) \left[
 e^{-\la/2} \left(3W + r \frac{dW}{dr}\right) + l(l+1) V
 - \frac{r^2}{2}H_0 - r^2K \right]
\non\\&
 - r \frac{d\eps}{dr} e^{-\la/2} W \:.
\label{e:deps}
\eal

To determine the amplitude of the GW strain,
which depends on the oscillation amplitudes,
it is necessary to use the total energy of oscillation for normalization.
In Newtonian approximation,
the kinetic energy per radial distance
of the $k$th eigenmode is \cite{Thorne69b}
\be\label{e:osenergy}
 \frac{dE_k}{dr} = \frac{\om^2}{2} (p+\eps) e^{(\la-\nu)/2}
 r^4 \left[ W^2 + l(l+1)V^2 \right] \:.
\ee
Since the potential energy is equal to
the kinetic energy during the oscillation \cite{Thorne67},
the total energy of the star's oscillation is $E_\text{os}=2E_k$.
In order to estimate the GW damping time
and compare it with the one from the full GR,
one can use the lowest-order post-Newtonian quadrupole formula
\cite{Thorne69b,Reisenegger92}
for the GW radiation power,
\be\label{e:pgw}
 p_\text{GW} = \frac{(l+1)(l+2)}{8\pi (l-1)l}
 \left[ \frac{4\pi \om^{l+1}}{(2l+1)!!}
 \int_0^R dr\, r^{l+2} \deps \right]^{2} \:.
\ee
To assess the accuracy of the quadrupole approximation,
considering the $f$-mode oscillation to
be efficient at extracting energy at the level of
oscillations and driving the emission of GWs, i.e.,
$E_\text{GW} \approx E_\text{os}$,
one can then compare the estimated damping time
\be\label{e:taues}
 \taues = \frac{E_\text{GW}}{p_\text{GW}} \:
\ee
with the damping time obtained from the full GR.

In this work, we use the real parts of
the fluid perturbation amplitude and frequency
in full GR to calculate the total energy of the oscillation
and the GW radiation power
using Eqs.~(\ref{e:osenergy}) and (\ref{e:pgw}), respectively,
and to estimate the damping time using Eq.~(\ref{e:taues}).
By comparing it with the accurate value from the full solution,
we can evaluate the approximation degree
of Eq.~(\ref{e:osenergy}) and (\ref{e:pgw}),
and finally, we estimate the amplitude of the GW strain.

\section{Numerical results}
\label{s:res}

\begin{figure}[t]
\vskip-4mm
\centerline{\includegraphics[width=0.5\textwidth]{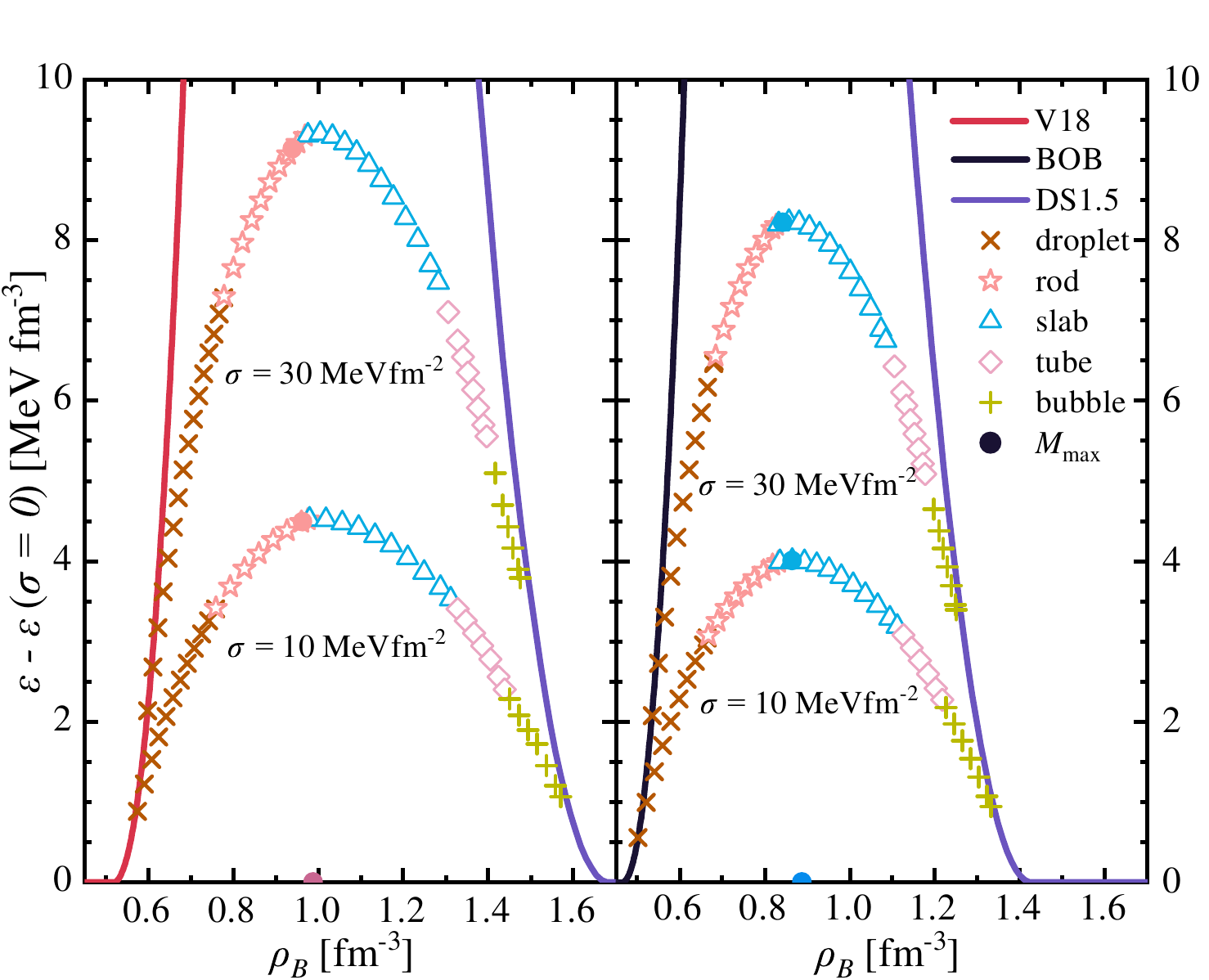}}
\vskip-3mm
\caption{
Energy density of the MP with $\sigma=10,30\mfms$
relative to the GC $(\sigma=0)$
as function of baryon number density,
obtained with two nucleonic EOSs, V18 (left panel) or BOB (right panel).
The maximum-mass configurations of HSs are indicated by full dots.
}
\label{f:eps-rho}
\end{figure}

\begin{figure}[t]
\vskip-2mm
\centerline{\includegraphics[width=0.5\textwidth]{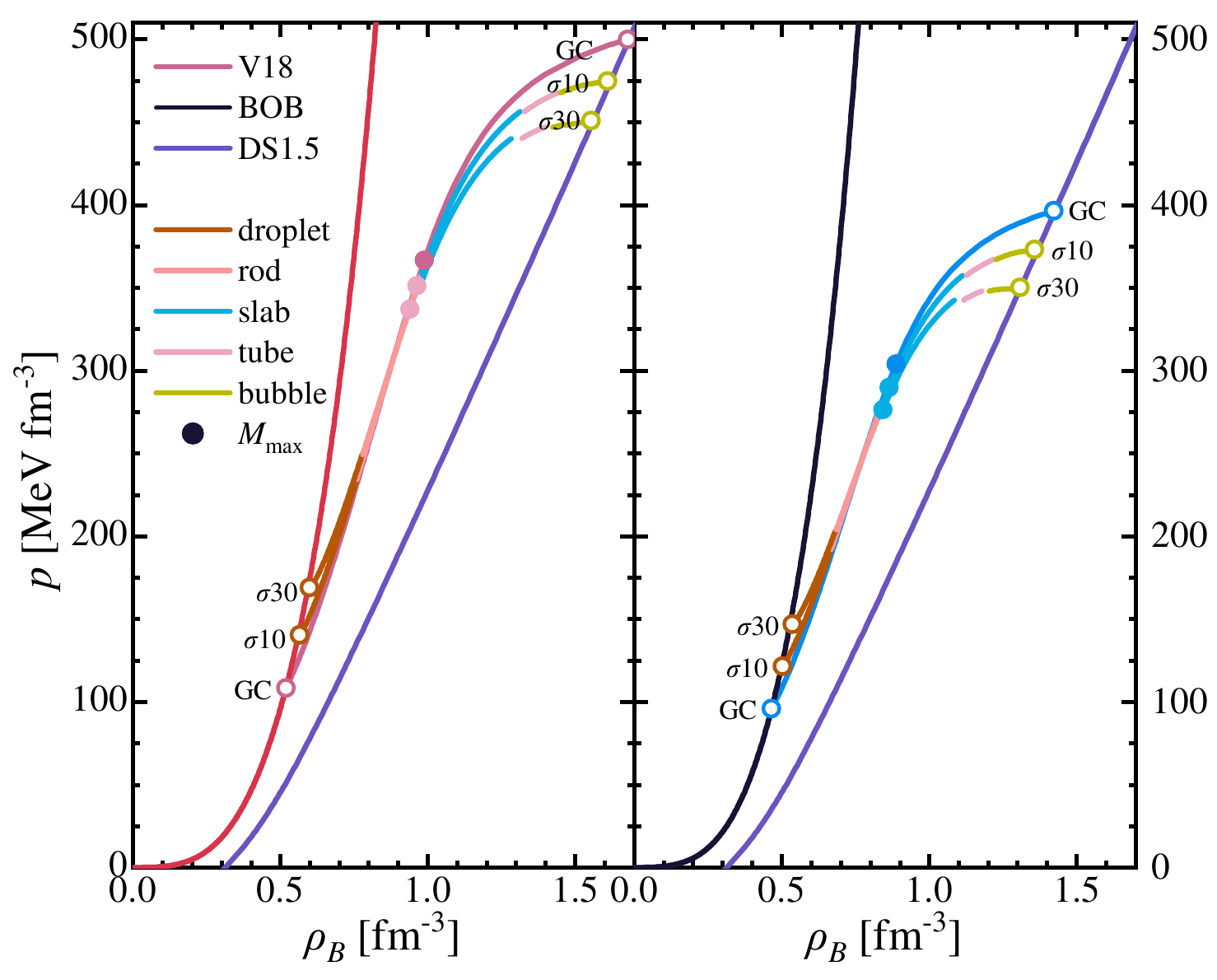}}
\vskip-4mm
\caption{
Same as Fig.~\ref{f:eps-rho}, for the
pressure as function of baryon number density.
}
\label{f:p-rho}
\end{figure}

\subsection{EOS and macroscopic features of neutron stars}

In this article,
we adopt the BHF EOS with BOB or V18 potential for NM.
Regarding the DSM for the QM EOS,
there is a free model parameter $\al_{\rm{DS}}$,
which represents the reduction of the in-medium effective interaction
\cite{Chen11}.
Here we choose $\al_{\rm{DS}}=1.5$ in combination with the BHF BOB/V18 EOSs,
labeled as BOB/V18+DS1.5 in the following,
which is compatible with present observations of NSs.
We choose $\sigma=10,30\mfms$
(denoted as ``$\sigma10,\sigma30$'')
to construct the pasta structures in the MP
as well as the GC (corresponding to $\sigma=0$) for comparison.
During the hadron-quark phase transition,
several pasta configurations may appear in a sequence of
QM droplet, rod, slab, tube, and bubble with increasing density.
In a Wigner-Seitz cell of the MP,
the distribution of quark and hadronic matter
is determined by minimizing the total energy
including surface and Coulomb contributions.

In Fig.~\ref{f:eps-rho}
we show the energy density of the MP with $\sigma=10,30\mfms$
relative to the GC $(\sigma=0)$
as function of baryon number density.
The results are calculated with V18 EOS (left panel)
or BOB EOS (right panel),
combined with QM EOS DS1.5.
For comparison, the energy densities of pure hadronic and pure quark phases
are also plotted.
In the MP, different symbols represent different geometrical structures.
At a given baryon number density around $0.5\fm3$,
when the energy density with quark droplets becomes lower than
that of pure hadronic matter,
those form within hadronic matter.
As the baryon number density increases,
other pasta configurations may appear,
and finally pure QM exhibits
lower energy density than pasta phases.
The points actually reached in the maximum-mass configurations of HSs
are indicated by full dots.
In all cases this occurs close to the rod/slab transition
(for V18/BOB),
near the maximum energy gain of the MP.
A pure QM phase is never reached.

\begin{figure}[t]
\vskip-3mm
\centerline{\includegraphics[width=0.5\textwidth]{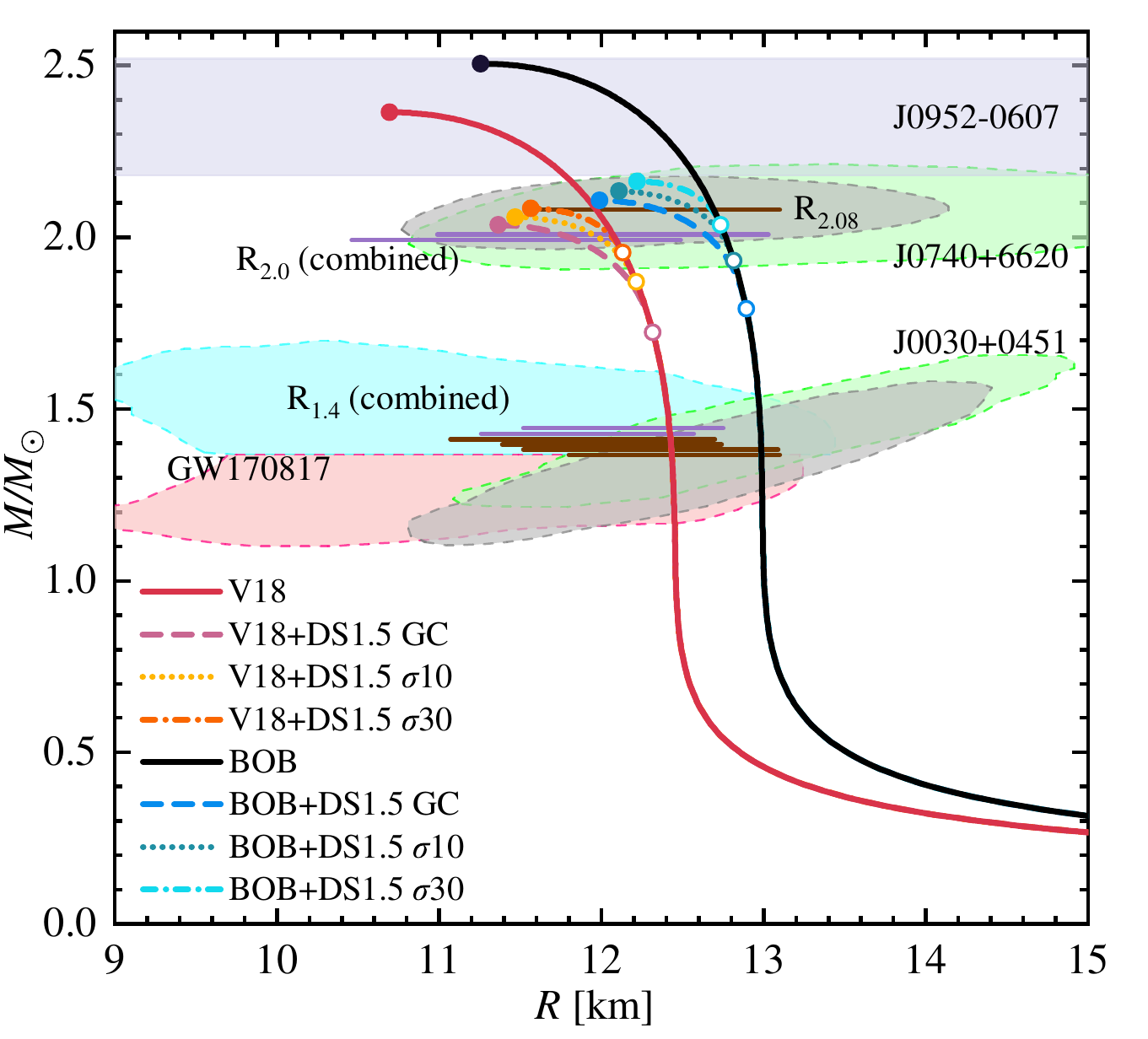}}
\vskip-3mm
\caption{
The mass-radius relations of NSs obtained with various EOSs.
Blank dots indicate the bifurcation points of pure NSs and HSs.
The horizontal bars indicate the limits on
$R_{2.08}$, $R_{2.0}$, and $R_{1.4}$
obtained in the combined NICER+GW170817 data analyses
of \cite{Miller21,Pang21,Raaijmakers21} (brown bars)
and the recent \cite{Rutherford24} (lilac bars).
The mass range of the heaviest currently known NS J0952-0607 is also shown.
The maximum-mass configurations are indicated by full dots.
}
\label{f:mr}
\end{figure}

As expected, with increasing $\sigma$
the density range of the MP shrinks
(eventually collapsing to a Maxwell construction).
This is seen more clearly in Fig.~\ref{f:p-rho},
where we show the pressure
$p=\rho_B^2 \partial(\eps/\rho_B)/\partial\rho_B$
as a function of the baryon number density
for hadronic, mixed, and quark phases.
with V18 and BOB EOS
and various values of $\sigma=0,10,30\mfms$.
The maximum-mass configurations are indicated by full dots.
It is clearly seen that up to that point
the pressures of pasta phases remain very close to those of the GC.
The main difference is a slightly higher onset density of the MP,
and a lower density and pressure in the center of maximum-mass HSs.
This corresponds to a larger maximum mass,
as seen in Fig.~\ref{f:mr}.

With the solutions of the TOV equations,
the mass-radius relations of pure NSs and HSs with the various EOSs
are shown in Fig.~\ref{f:mr}.
The broken curve segments indicate the hybrid star branches
with various values of surface tension $\sigma$.
All EOSs considered in the article are compatible
with the two-solar-mass constraint
\cite{Arzoumanian18,Antoniadis13,Fonseca21}.
However, the reduced maximum masses of some hybrid model might conflict with
the recent data $M=2.35\pm0.17\ms$ for PSR J0952-0607 \cite{Romani22}
or provide an additional constraint on the surface tension.
The mass-radius results of the NICER mission for the pulsars
J0030+0451 \cite{Riley19,Miller19,Vinciguerra24}, and
J0740+6620 \cite{Riley21,Miller21,Pang21,Raaijmakers21,Salmi24,Dittmann24},
PSR J0437-4715 \cite{Choudhury24}
are also plotted in the figure.
The combined analysis of those pulsars together with GW170817
yields improved limits on
$R_{2.08}$ \cite{Miller21},
$R_{2.0}$ \cite{Rutherford24}, and
$R_{1.4}$ \cite{Pang21,Raaijmakers21,Rutherford24},
which are shown as horizontal bars.
The V18 is clearly preferred as nucleonic EOS,
also in terms of its maximum mass.
We will not delve further into the mass and radius of NSs here
and refer to recent literature \cite{Wei19,Sun21,Wei21,Burgio21}
for more details.

\begin{figure}[t]
\vskip-5mm
\centerline{\includegraphics[width=0.5\textwidth]{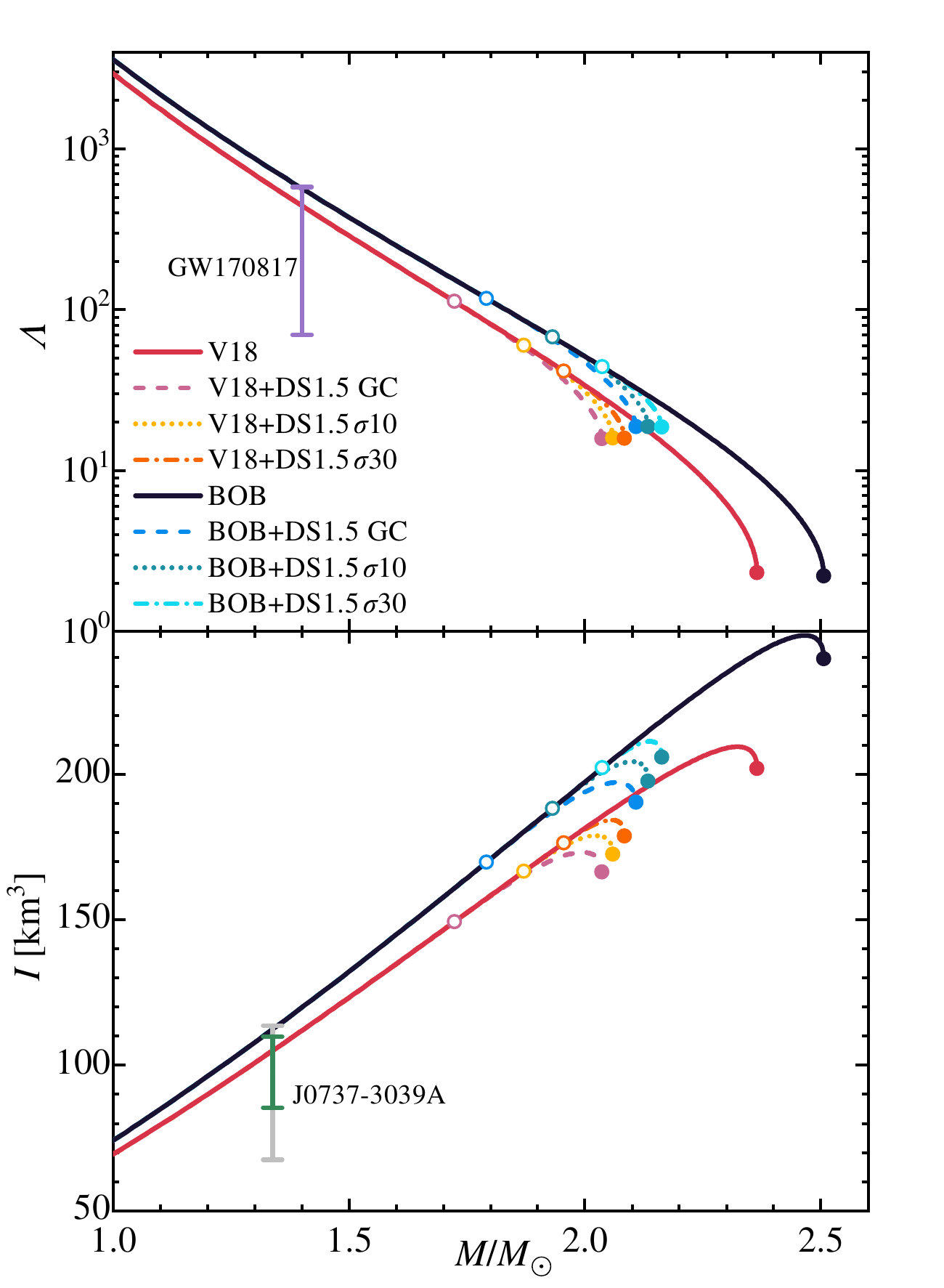}}
\vskip-4mm
\caption{
The dimensionless tidal deformability (upper panel) and the MOI (lower panel)
of NSs vs NS mass $M$ for various EOSs.
Observational values for GW170817 and the pulsar J0737-3039A are shown,
see text.
}
\label{f:tdmoi-m}
\end{figure}

\begin{figure}[t]
\vskip-2mm
\centerline{\includegraphics[width=0.5\textwidth]{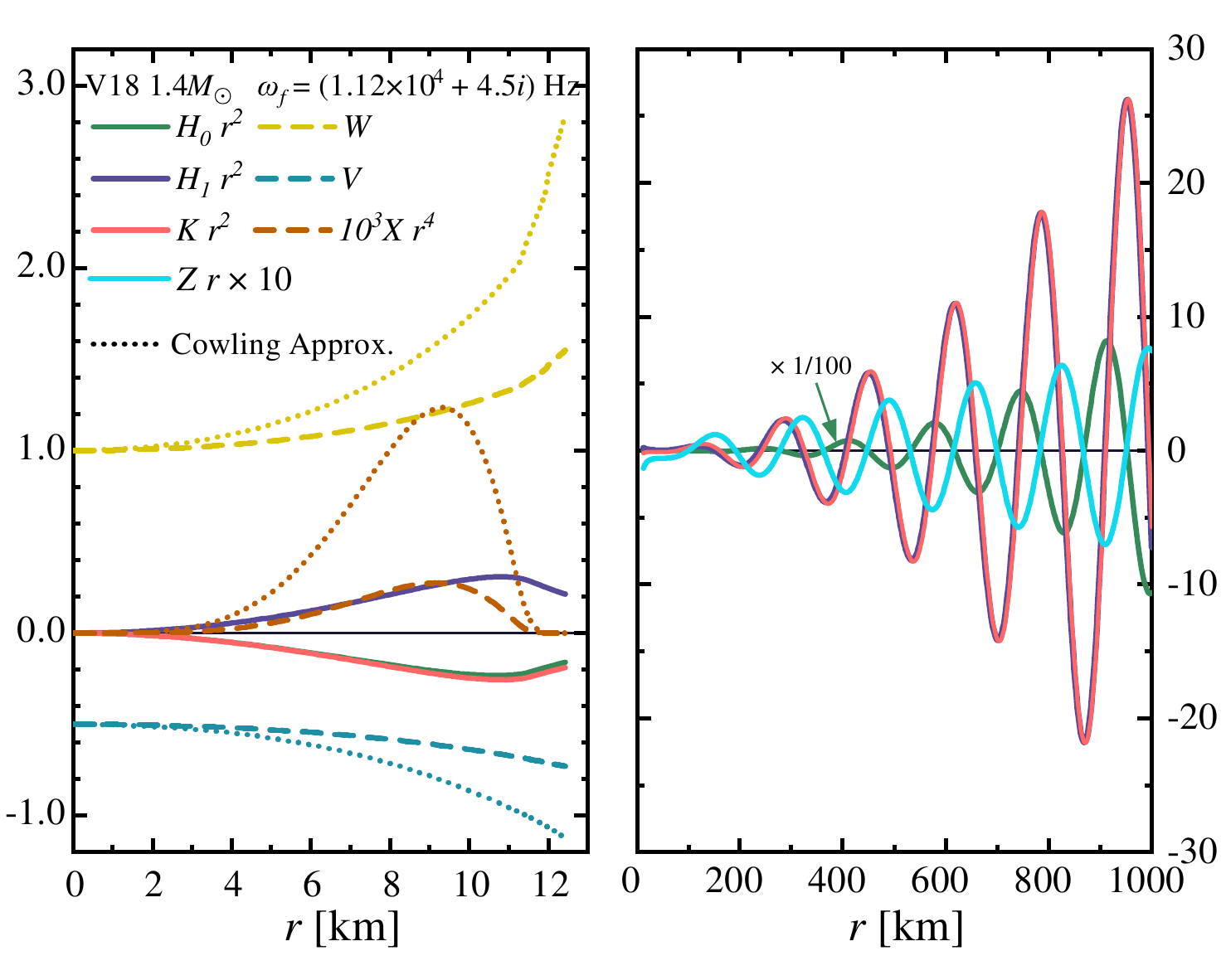}}
\vskip-5mm
\caption{
Real parts of the dimensionless metric perturbation amplitudes $H_0,K,H_1,Z$
and fluid perturbation amplitudes $W,V,X$
for a nonradial ($l=2$) $f$-mode oscillation,
inside (left panel) and outside (right panel) a $1.4\ms$ NS.
Results are obtained with the V18 EOS.
}
\label{f:internal}
\end{figure}

\OFF{
PSR J1614-2230 ($M = 1.908\pm0.016M_\odot$) \cite{Arzoumanian18},
PSR J0348+0432 ($M = 2.01\pm0.04M_\odot$) \cite{Antoniadis13}, and
PSR J0740+6620  2.14-0.09+0.10 \cite{Cromartie20}, OLD!
PSR J0740+6620 ($M = 2.08\pm0.07M_\odot$), \cite{Fonseca21},
$M=2.35\pm0.17\ms$ for PSR J0952-0607 \cite{Romani22}

We also mention the combined estimates of the mass and radius
of the isolated pulsar PSR J0030+0451 observed recently by NICER,
$M=1.44^{+0.15}_{-0.14}\ms$ and $R=13.02^{+1.24}_{-1.06}\,$km
\cite{Miller19,Riley19}, or
$M=1.36^{+0.15}_{-0.16}\ms$ and $R=12.71^{+1.14}_{-1.19}\,$km
\cite{Miller21,Riley21},

and for PSR J0740+6620 with
$R(2.08\pm0.07M_\odot) = 13.7^{+2.6}_{-1.5}$ km
\cite{Riley21} and
$R(2.072^{+0.067}_{-0.066}M_\odot) = 12.39^{+1.30}_{-0.98}$ km
\cite{Miller21}.
\cite{Riley21,Miller21,Pang21,Raaijmakers21}

and in particular the result of the combined GW170817+NICER analysis
$R_{2.08}=12.35\pm0.75\,$km \cite{Miller21}, and
$R_{1.4}=
$12.45\pm0.65$ km \cite{Miller21}, 
$11.94^{+0.76}_{-0.87}$ km \cite{Pang21}, and
$12.33^{+0.76}_{-0.81}$ km or
$12.18^{+0.56}_{-0.79}$ km \cite{Raaijmakers21}
$12.28^{+0.50}_{-0.76}$ km \cite{Rutherford24}PP
$12.01^{+0.56}_{-0.75}$ km \cite{Rutherford24}CS
$R_{2.0}=12.33^{+0.70}_{-1.34}$ km \cite{Rutherford24}PP
$R_{2.0}=11.55^{+0.94}_{-1.09}$ km \cite{Rutherford24}CS
}

Rather, in Fig.~\ref{f:tdmoi-m}
we show the dimensionless tidal deformability $\Lambda$, Eq.~(\ref{e:tdef}),
(upper panel),
and the MOI $I$, Eq.~(\ref{e:moi}),
(lower panel) as functions of the NS mass.
The constraint $\Lambda_{1.4}=190^{+390}_{-120}$
obtained by the analysis of the observational data
from GW170817 and its electromagnetic counterpart
\cite{Abbott18} (purple bar),
as well as the constraints on the MOI
of the pulsar J0737-3039A   
based on the GW170817 tidal measurement
\cite{Landry18} (grey bar)
or by using the multimessenger constraints on the radius \cite{Dietrich20}
in combination with the radius-MOI relation \cite{Lattimer19} (green bar)
are also shown in the figure.
These constraints are all compatible with our pure NS results.
HSs appear only with much higher masses
$M\gtrsim1.7\ms$
and their tidal deformability and MOI
are smaller than those of pure NSs.

\begin{figure}[t]
\vskip-2mm
\centerline{\includegraphics[width=0.5\textwidth]{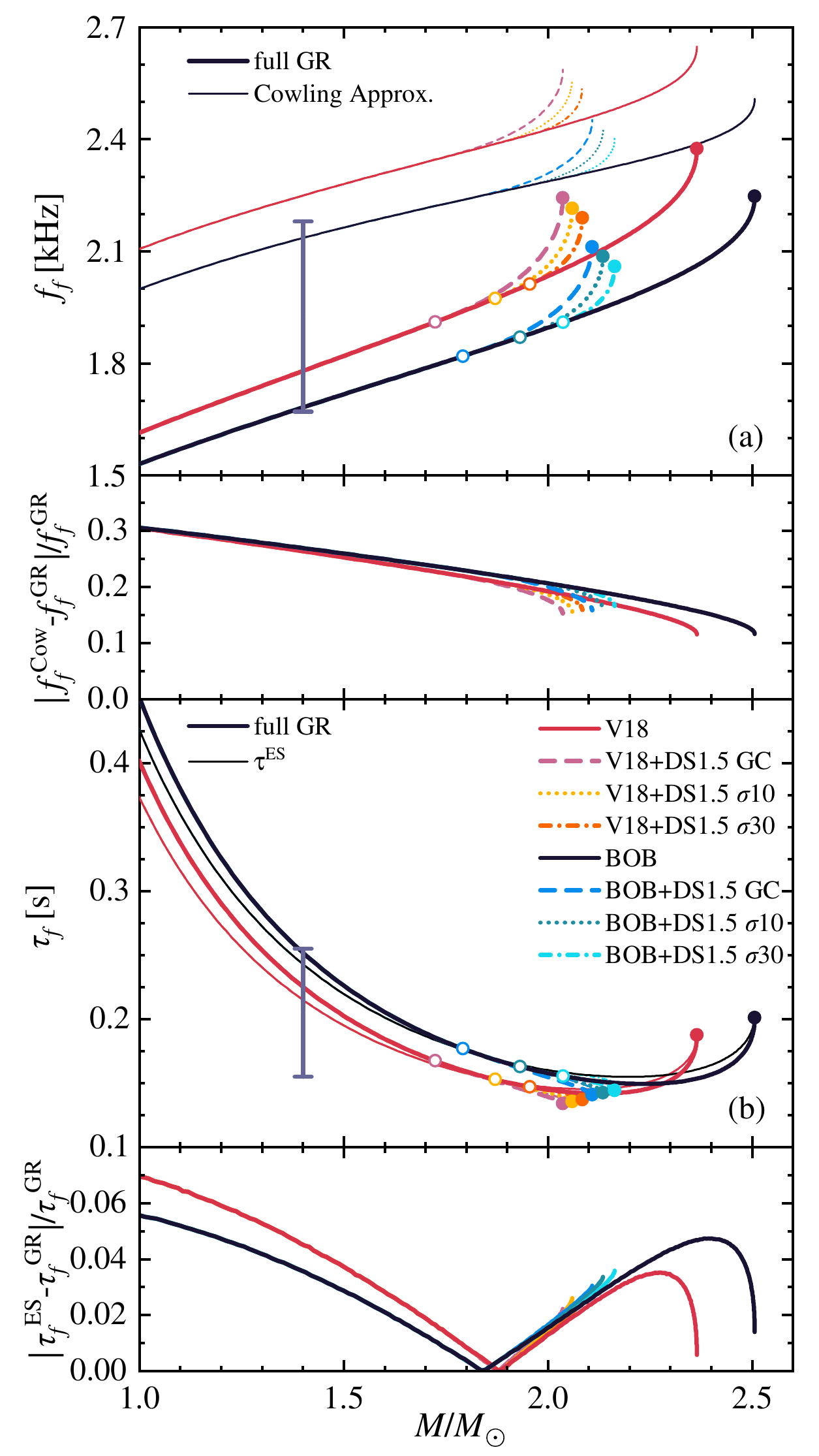}}
\vskip-4mm
\caption{
(a) GW Frequencies and (b) damping times
of $(l=2)\ f$-mode oscillations vs NS mass $M$ with various EOSs.
The vertical bars indicate the GW170817 constraint \cite{Wen19}.
The lower parts of the two panels show relative deviations
of frequencies in the Cowling approximation
and damping times estimated by Eq.~(\ref{e:taues})
from the full solutions, respectively.
}
\label{f:ftau-m}
\end{figure}

\begin{figure}[t]
\vskip-1mm
\centerline{\includegraphics[width=0.5\textwidth]{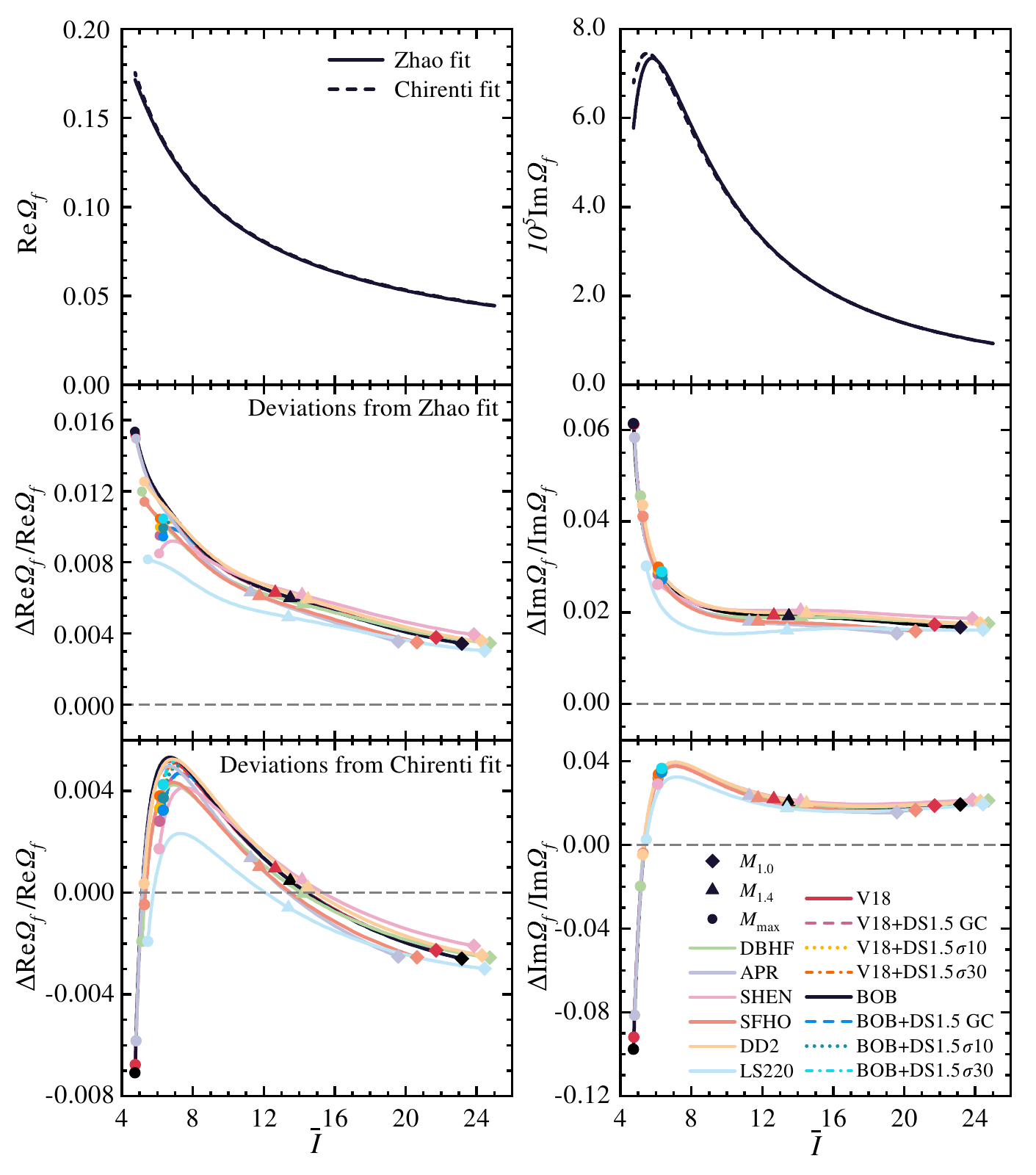}}
\vskip-2mm
\caption{
Top panels:
The URs of \cite{Zhao22b} (Zhao) and \cite{Chirenti15} (Chirenti)
between dimensionless $f$-mode frequency $\Omega_f$
(real and imaginary part)
and dimensionless MOI $\bar{I}$,
and their relative deviations from our considered EOSs
(lower panels).
}
\label{f:mf-moi/m3}
\end{figure}

\begin{figure}[t]
\vskip-1mm
\centerline{\includegraphics[width=0.5\textwidth]{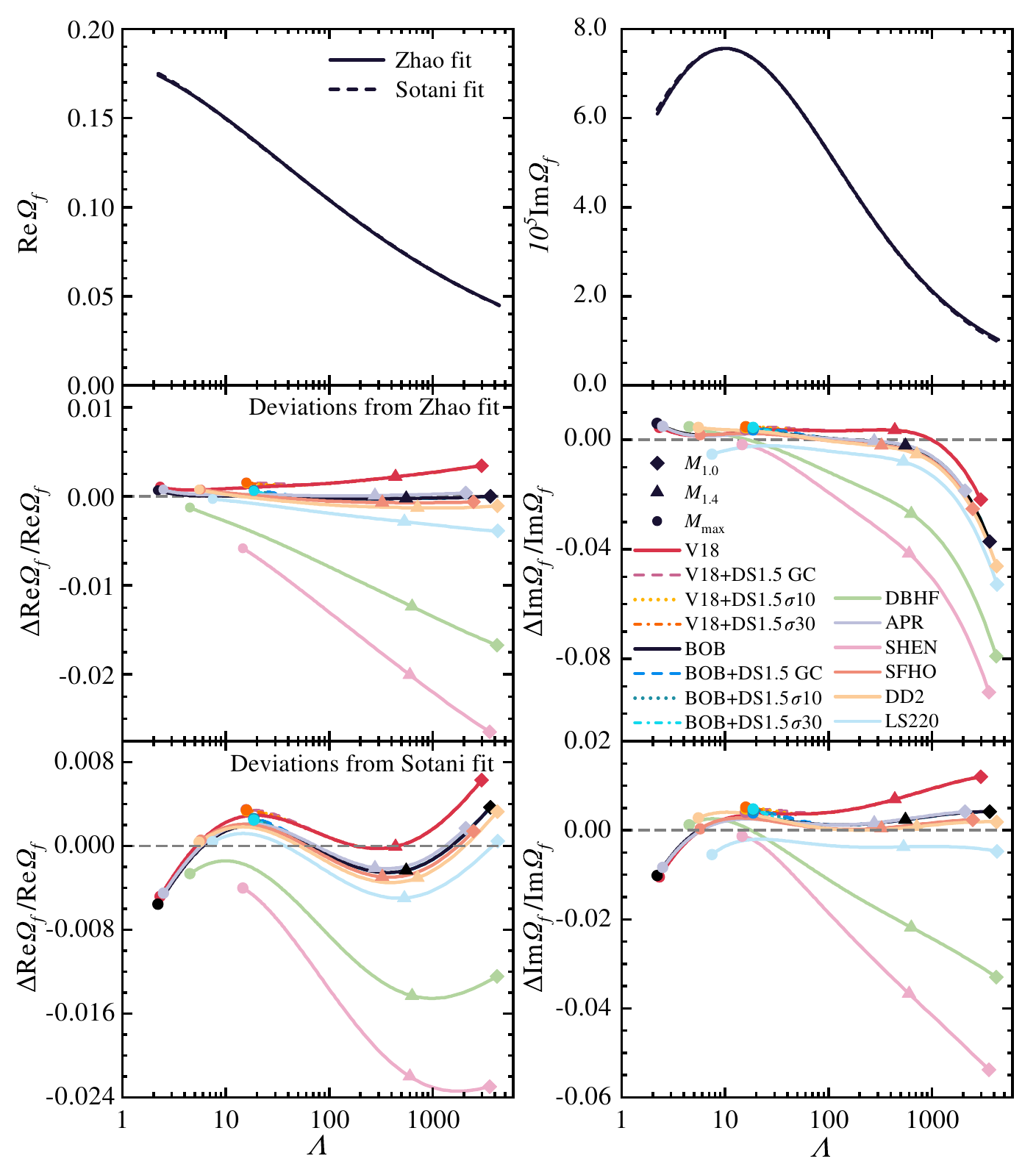}}
\vskip-2mm
\caption{
Same as Fig.~\ref{f:mf-moi/m3},
for the $\Omega_f(\Lambda)$ relation,
in comparison with URs of \cite{Zhao22b} (Zhao) and \cite{Sotani21} (Sotani).
}
\label{f:mf-lmbda}
\end{figure}

\begin{figure}[t]
\vskip-1mm
\centerline{\includegraphics[width=0.5\textwidth]{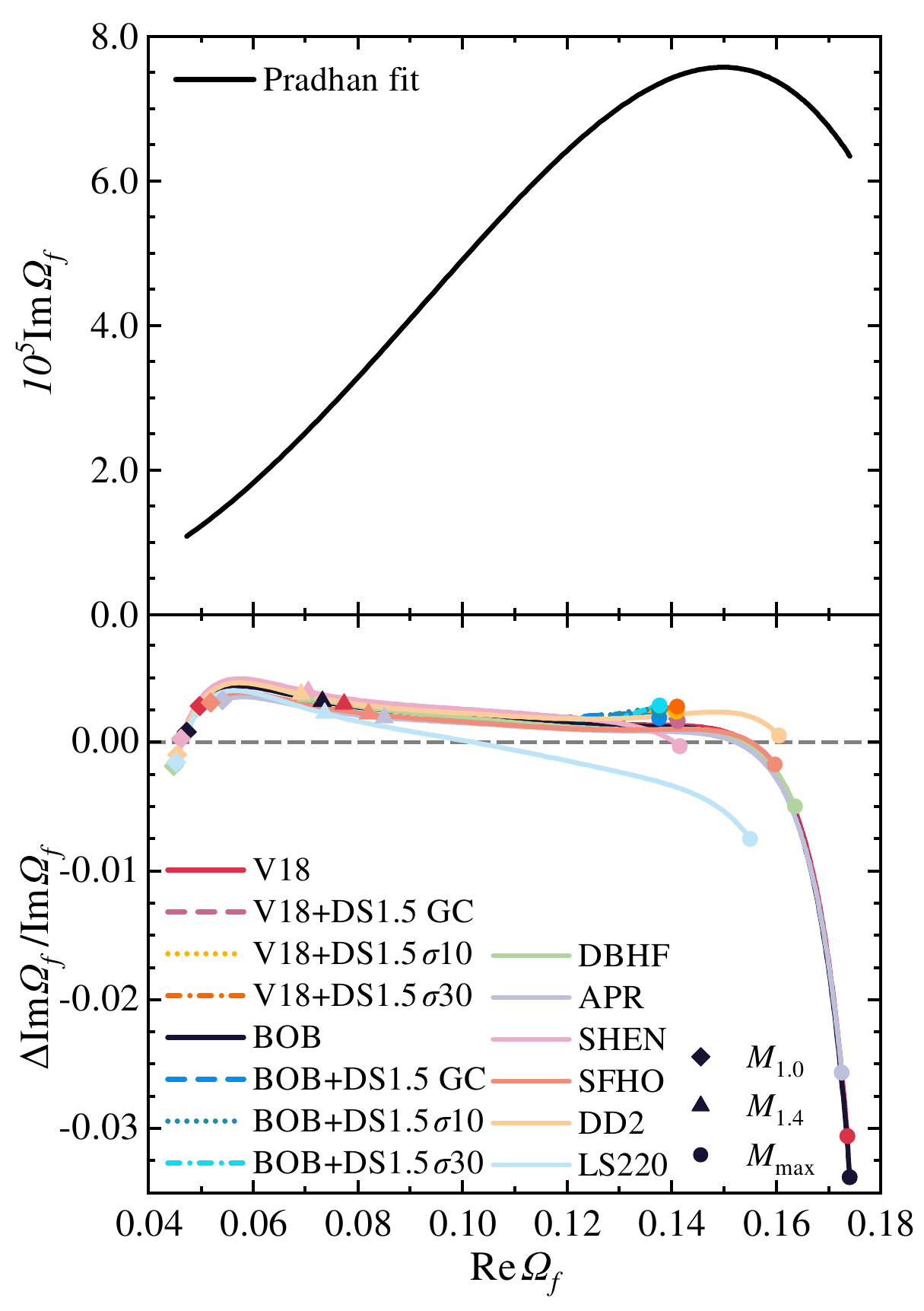}}
\vskip-4mm
\caption{
Upper panel :
The UR of \cite{Pradhan22} (Pradhan)
between imaginary part and real part
of dimensionless $f$-mode frequency $\Omega_f$,
and relative deviations from our considered EOSs
(lower panel).
}
\label{f:mf-mtau}
\end{figure}

\begin{figure}[t]
\vskip-1mm
\centerline{\includegraphics[width=0.5\textwidth]{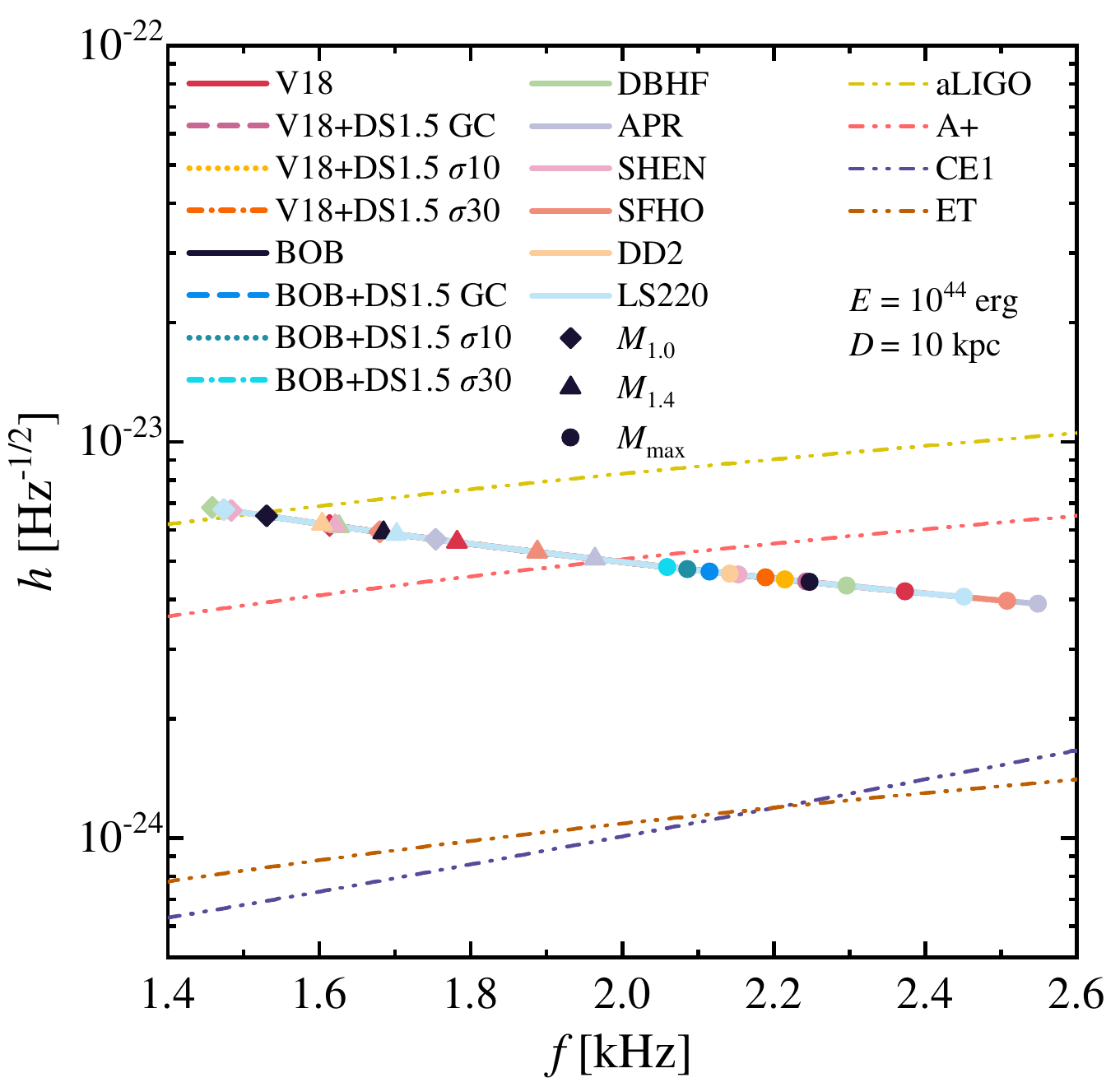}}
\vskip-3mm
\caption{
The peak strain $h$, Eq.~(\ref{e:h2}), of a burst of GWs
emitted at frequency $f$ from a $f$-mode oscillation.
The double-dotted lines are sensitivities of
Advanced LIGO (aLIGO), A+, Cosmic Explorer (CE1),
and Einstein Telescope (ET).
}
\label{f:h-f}
\end{figure}

\subsection{$f$-mode oscillations}

In this work, we investigate the
quadrupole ($l=2$) $f$-mode oscillations of both pure NSs and HSs
with pasta-construction phase transition,
in the frame of full GR.
For illustration,
we show in Fig.~\ref{f:internal} a typical solution for the real parts of the
dimensionless metric ($H_0,K,H_1,Z$)
and fluid ($W,V,X$)
perturbation amplitudes
inside (left panel) and outside (right panel) a $1.4\ms$ NS
computed with the V18 EOS.
For a canonical $f$-mode,
there are no nodes for both fluid and metric perturbation amplitudes
inside the star.
The dotted curves represent the fluid perturbation amplitudes
inside the star under the Cowling approximation \cite{Cowling_41},
in which the metric perturbations are usually disregarded.
The fluid perturbation amplitudes $W$ and $V$ in both cases
increase with the radius $r$,
and this indicates that the oscillation occurs mainly in the crust,
thus the $f$-mode oscillation hardly bears the properties of the core.
However, one can see that the fluid perturbations
under the Cowling approximation
are more intense compared to the full GR case,
where the fluid transfers oscillation energy to the metric perturbation,
which in turn suppresses the fluid perturbation.
Therefore,
where the metric perturbation amplitudes are significant,
the suppression of fluid perturbations is stronger,
and this is clearly displayed in the figure.

The corresponding $f$-mode frequency is
$\om_f = (1.12\times10^4 + 4.46i)\,$Hz.
In comparison, our previous work \cite{Zheng23}
with the Cowling approximation gives
$\om_f = 1.41\times10^4\,$Hz,
about $25\%$ larger.
Thus, it is necessary to investigate the $f$-mode oscillations
in the frame of full GR.
Note that $H_0$ and $K$ inside the star
and $H_1$ and $K$ outside the star are almost identical.
These results are similar to those in literature \cite{Zhao22b}.

Now we turn to a comparison of the $f$-mode frequencies in pure NSs and HSs.
In Fig.~\ref{f:ftau-m},
we show the $f$-mode frequencies (a) and damping times (b)
of pure NSs and HSs as functions of their masses.
We also show the constraints (vertical bars) for a canonical NS
obtained by using the deduced GW170817 tidal deformability
\cite{Wen19,Pratten20}.

In full GR,
the frequency of both pure NSs and HSs lies in the range $1.5\text{--}2.5\khz$,
and increases with the NS mass.
One can see that for a given mass,
the frequency for a HS is larger than for a pure NS.
Unfortunately, the difference is not larger than
the one caused by different models of pure NSs.
It means that the $f$-mode frequency is mainly influenced
by the global features of the star \cite{Andersson98,Kokkotas99},
but insensitive to internal components.
For comparison,
we also show the results obtained using the Cowling approximation (thin curves)
and their relative deviations (lower part) from the full solutions,
which decrease with increasing mass,
from $30\%$ to $10\%$.
The obtained trend of improving accuracy with increasing mass
is in good agreement with previous results
\cite{Chirenti15,Pradhan22,Kunjipurayil22}.

In the same figure, one can see that the damping times in full GR
are of the order of a few tenths of a second
and are almost indistinguishable between NSs and HSs.
Compared to the  $g_1$-mode damping time of about
$10^4\text{--}10^8\,$s \cite{Zheng23},
that of the $f$-mode is several orders of magnitude shorter,
which means stronger GW radiation power,
and thus more likely to be detected.
Ref.~\cite{Kokkotas01c} showed that
$f$-mode frequencies could be accurately deduced from detected signals
with improved sensitivity of GW detectors,
whereas the damping time could not.
However, as discussed in the following,
URs can not only be used to extract the global properties of NSs,
but also be helpful in constraining the $f$-mode damping time.

In Fig.~\ref{f:ftau-m}(b),
we also show the estimated damping times $\taues$ (thin lines),
Eq.~(\ref{e:taues}),
obtained using the lowest-order post-Newtonian quadrupole formula,
and their relative deviations from the full solutions (lower part).
One can see that the estimated values deviate less than $7\%$
from the full solution,
which demonstrates the accuracy of the quadrupole approximation,
Eqs.~(\ref{e:osenergy}) and (\ref{e:pgw}).

\subsection{Universal relations}

The technique of inferring NS features from nonradial oscillations
was first introduced by Andersson and Kokkotas \cite{Kokkotas99,Andersson98}.
They evidenced the URs between
dimensionless frequency or dimensionless damping time
and mean mass density or compactness parameter $\beta=M/R$ of the star.
Subsequently \cite{Lau09,Chirenti15},
a much more precise EOS-insensitive relation between
dimensionless frequency $\Omega_f \equiv M\om_f$
and dimensionless MOI $\bar{I} \equiv I/M^3$
was found.
Recently, this UR was confirmed by Ref.~\cite{Zhao22b}
for all three stellar categories, hadronic, hybrid with MC, and quark.
Furthermore,
as analyzed in \cite{Chan14,Sotani21,Zhao22b},
the $\Omega_f(\bar{I})$ UR implies a related $\Omega_f(\Lambda)$ UR,
since the EOS-insensitive $\bar{I}-\Lambda-Q$ UR is well known \cite{Yagi13}.
Both types of URs are expressed in terms of polynomial fits,
\bal
 \Omega_f &= \sum_{i=0}^7 a_i \bar{I}^{-i/2} \;,
\label{e:fit1}
\\
 \Omega_f &= \sum_{i=0}^7 b_i (\ln\Lambda)^i \;,
\label{e:fit2}
\eal
where $a_1$ and $b_i$ are complex parameters.
Furthermore, the URs for real and imaginary parts of $\Omega_f$
imply a UR between both quantities \cite{Pradhan22}.


We will now check the compliance of our nucleonic and pasta-phase hybrid EOSs
with these URs.
The results are displayed in Figs.~\ref{f:mf-moi/m3},
\ref{f:mf-lmbda}, and \ref{f:mf-mtau}
for $\Omega_f(\bar{I})$, $\Omega_f(\Lambda)$,
and $\immf(\remf)$, respectively.
The upper panels show the URs given by
Zhao \cite{Zhao22b}, Chirenti \cite{Chirenti15}
and
Zhao \cite{Zhao22b}, Sotani \cite{Sotani21}
and
Pradhan \cite{Pradhan22}, respectively, whereas
the middle and bottom panels show the relative deviations
$(\Omega_\text{EOS}-\Omega_\text{UR})/\Omega_\text{UR}$
of our various EOSs from the fits.
For comparison,
we also show results with the nucleonic
DBHF \cite{Gross-Boelting99},
APR \cite{Akmal98},
Shen2020 \cite{Shen20},
SFHO \cite{Hempel10,Steiner13},
DD2 \cite{Typel10,Grill14},
and LS220 \cite{Lattimer91,Oertel12} EOS.
The figures indicate that the URs
for the real (imaginary) parts of $\Omega_f$
are valid within about
1\% (5\%) for $\Omega_f(\bar{I})$,
3\% (10\%) for $\Omega_f(\Lambda)$
and 3\% for $\immf(\remf)$.
The differences between different nucleonic EOSs are much larger than
the deviations of hybrid branches.
Therefore, these URs cannot give any indications for the presence of
QM in NSs.

It is believed that $f$-mode frequencies could be accurately deduced
from detected signals
with improved sensitivity of future GW detectors \cite{Kokkotas01c},
whereas the damping time may not be detected with such good accuracy.
With these URs and more precise constraints on the tidal deformability of NSs
in the future,
the $f$-mode frequency and damping time can be constrained
into a more precise range \cite{Wen19}.
Besides that, the tidal deformability
inferred during the premerger inspiral phase
along with the $f_\text{peak}$
(caused by fundamental quadrupole oscillation)
detected during the postmerger phase can also be used
to verify the presence of quarks in the interior of NSs \cite{Blacker20}.

In addition, measurements are planned for the MOI
of PSR J0737-3039~A \cite{Lyne04},
the $1.338\ms$ primary component of
the first double-pulsar system PSR J0737-3039 \cite{Burgay03},
based on the long-term pulsar timing to determine the periastron advance
due to relativistic spin-orbit coupling \cite{Damour98}.
The MOI of PSR J0737-3039~A is expected to be measured
with $\sim10\%$ accuracy within the next decade,
which will provide complementary constraints on the EOS
and more precise limitations on the $f$-mode characteristics.

\subsection{Prospects of observation}

In the previous section,
we examined the compliance of our nucleonic
and pasta-phase hybrid EOSs with several URs.
In the following, we investigate a closely related observable,
the GW peak strain of the oscillations,
\be
 h = \sqrt{\tau^{ES}}~h_+
 = \frac{\sqrt{15E_\text{GW}}}{2\pi D f} \:,
\label{e:h2}
\ee
where $\tau^{ES}$ given in Eq.~(\ref{e:taues})
is the estimated damping time in quadrupole approximation,
and $h_+$ is given in Eq.~(\ref{e:hp1}).
The last expression is obtained
after some mathematical transformations, i.e.,
using Eqs.~(\ref{e:pgw}) and (\ref{e:taues})
to eliminate the integral term in Eq.~(\ref{e:Q33}).
It shows independence from the EOS,
only depending on frequency $f$ and energy $E_\text{GW}$ of the GW,
and the source distance $D$.
This expression is consistent with Eq.~(3) in \cite{Ho20},
apart from an extra factor $\sqrt{3}/2$ here.
We suspect this is due to ignorance of cross polarization
or the inclination angle in Eq.~(\ref{e:hp1}).

As the oscillation occurs mainly in the crust
(see Fig.~\ref{f:internal}),
it may be strongly excited by the glitching behavior of an isolated star
\cite{Sidery10,Ho20,Haskell24}.
One of the mechanisms is the direct excitation of the $f$ mode,
following a glitch \cite{Ho20}.
Another possibility is that the glitch is triggered
by a fracture in the solid crust.
Such a starquake could excite multiple modes of oscillation,
including $f$ mode and its overtones \cite{Keer15}.
In our best-case scenario,
we assume that the glitch energy is entirely emitted in GWs
via $f$-mode oscillation.
The energy of the glitch can be estimated by combining the MOI,
the pulsar spin frequency and the glitch size.
The estimated energies of observed pulsar glitches are
in the range of $10^{37}\text{--}10^{44}\;$erg,
and the majority of known pulsars and glitching pulsars
are at distances $D<6\;$kpc \cite{Ho20}.
Note that most glitching pulsars have relatively low spin frequencies,
about $1\text{--}30\;$Hz \cite{Ho20},
thus it is reasonable to adopt the non-rotating approximation
for the $f$-mode oscillation.

Choosing the maximum energy scale $E \sim 10^{44}\;$erg
and a typical distance $D\sim10\;$kpc (star in our galaxy),
we show in Fig.~\ref{f:h-f}
the corresponding GW peak strain $h$ of $f$-mode oscillations
obtained from various considered EOSs
against the sensitivity curves of
Advanced LIGO (aLIGO) \cite{Abbott18b},
A+ \cite{A+},
Cosmic Explorer (CE1) \cite{CE&ET},
and Einstein Telescope (ET) \cite{Hild11}.
The GW peak strain from NSs in our galaxy is above $10^{-24}$
across the entire $f$-mode frequency range,
falling within the detection capability of next-generation GW detectors.
In the low-frequency region
($\lesssim 2\;$kHz),
the GW peak strain can even reach the detection threshold
of current GW detectors.
For nearer glitching pulsars,
the frequency range of a GW signal detectable by current GW detectors
will increase.

Besides by glitches,
the $f$-mode oscillation might also be triggered
during a supernova explosion \cite{Radice19}
or during the post-merger phase of a binary NS merger
\cite{Kokkotas01,Stergioulas11,Vretinaris20,Soultanis22},
which may increase the energy of GWs,
resulting in stronger GW strain amplitudes
and enhancing their detectability.

\section{Conclusions}
\label{s:end}

With the next generation of GW telescopes,
the GWs from quasinormal modes of NS oscillation may be detectable,
as we have also shown in this work,
where we investigated non-radial quadrupole $f$-mode oscillations
of cold and isolated NSs,
including pure NSs and HSs.
We adopted the BHF theory for NM, the DSM for QM,
and the pasta construction for their phase transition.
Based on the equilibrium structure,
we solved the oscillation equations within full GR,
and obtained the metric and fluid-displacement perturbations
inside and outside NSs
as well as the complex eigenfrequencies for the various EOSs.

If a moderate hadron-quark surface tension $\sigma$ is employed,
in the MP several pasta configurations may appear in sequence of
droplet, rod, slab, tube, and bubble with increasing density.
As $\sigma$ increases,
the MP density range shrinks.
The pasta phase also influences the mass-radius relations of HSs.
All EOSs considered in the article are compatible with
two-solar-mass observations
and NICER mass-radius constraints of NSs.
The corresponding tidal deformability and MOI
also satisfy the constraints from the analysis of GW170817.
The $f_f$ of both pure NSs and HSs lies in the range $1.5\text{--}2.5\khz$,
and the damping times are of the order of a few tenths of a second
with a corresponding GW radiation power
that might be detectable.
However, the differences of $f$-mode characteristics
between HSs and pure NSs are not large,
compared to the differences between pure NSs with different models.

Previous works have studied $f$-mode frequencies and damping times,
and proposed EOS-insensitive URs with dimensionless MOI $\bar{I}$
and dimensionless tidal deformability $\Lambda$.
The validity for hadronic, pure quark, and hybrid stars was confirmed.
However, HSs were studied only with MC phase transition,
GC phase transition, or crossover.
We extended these studies to HSs with the pasta-construction phase transition,
showing that the URs are also valid in this case,
which means that they cannot be used to indicate
the presence of QM in NSs.
The URs were found to be valid within a few percents
for the EOSs we tested.

Apart from cold isolated NSs,
quadrupole oscillations also occur in various newly born NSs,
after NS mergers or supernova explosions,
which are expected to be more energetic and easier observable.
In these cases,
one needs to consider more realistic environment effects,
such as the EOS at finite temperature,
the temperature/entropy distribution in NSs,
the neutrino trapping effects,
viscosity and heat conduction,
and also (differential) rotation.
We leave these issues to future work.

\begin{acknowledgments}

We would like to thank Tianqi Zhao for helpful and selfless discussions
of the numerical treatment.
Zi-Yue Zheng and Xiao-Ping Zheng are supported by
the National Natural Science Foundation of China (Grant No.~12033001)
and
the National SKA Program of China (Grant No.~2020SKA0120300).
Jin-Biao Wei and Huan Chen acknowledge financial support from the
National Natural Science Foundation of China (Grant No.~12205260).

\end{acknowledgments}

\newcommand{\epja}{Euro. Phys. J. A}
\newcommand{\aap}{Astron. Astrophys.}
\newcommand{\apjl}{Astrophys. J. Lett.}
\def\jcap{Journal of Cosmology and Astroparticle Physics}
\def\jcap{JCAP}
\newcommand{\mnras}{Mon. Not. R. Astron. Soc.}
\newcommand{\nphysa}{Nucl. Phys. A}
\newcommand{\physrep}{Phys. Rep.}
\newcommand{\plb}{Phys. Lett. B}
\newcommand{\ppnp}{Prog. Part. Nucl. Phys.}
\bibliographystyle{apsrev4-1}
\bibliography{pasta}

\end{document}